\documentclass[twocolumn,secnumarabic,amssymb, nobibnotes, aps, prx, superscriptaddress]{revtex4-2}
\usepackage{lipsum}
\usepackage{xcolor}
\usepackage{changes}
\usepackage{graphicx}
\graphicspath{{./figures/}}     %

\usepackage{siunitx}
\usepackage{tikz}
 
\usepackage{ulem}
\usepackage{natbib}
\usepackage{mathtools}
\usepackage{cleveref}
\usepackage{xspace}
\usepackage{braket}
\usepackage{comment}

\newcommand{\fm}[1]{\ifmmode#1\else$#1$\fi}

\newcommand{\Ti}{\fm{^{48}\text{Ti}^{+}}\xspace}
\newcommand{\Ca}{\fm{^{40}\text{Ca}^{+}}\xspace}
\newcommand{\Al}{\fm{^{27}\text{Al}^{+}}\xspace}

\newcommand{\Fa}{a$^{4}$F\xspace}

\begin{document}
\bibliographystyle{plainnat}

\title{%
 Quantum logic control of a transition metal ion\\
 }

\author{Till Rehmert}%
\email{till.rehmert@ptb.de}
\affiliation{Physikalisch-Technische Bundesanstalt, Bundesallee 100, 38116 Braunschweig, Germany}
\affiliation{Institut für Quantenoptik, Leibniz Universität Hannover, Welfengarten 1, 30167 Hannover}
\author{Maximilian J. Zawierucha}%
\email{maximilian.zawierucha@ptb.de}
\affiliation{Physikalisch-Technische Bundesanstalt, Bundesallee 100, 38116 Braunschweig, Germany}
\affiliation{Institut für Quantenoptik, Leibniz Universität Hannover, Welfengarten 1, 30167 Hannover}
\author{Kai Dietze}
\affiliation{Physikalisch-Technische Bundesanstalt, Bundesallee 100, 38116 Braunschweig, Germany}
\affiliation{Institut für Quantenoptik, Leibniz Universität Hannover, Welfengarten 1, 30167 Hannover}
\author{Piet O. Schmidt}%
\email{piet.schmidt@quantummetrology.de}
\affiliation{Physikalisch-Technische Bundesanstalt, Bundesallee 100, 38116 Braunschweig, Germany}
\affiliation{Institut für Quantenoptik, Leibniz Universität Hannover, Welfengarten 1, 30167 Hannover}
\author{Fabian Wolf}%
\email{fabian.wolf@ptb.de}
\affiliation{Physikalisch-Technische Bundesanstalt, Bundesallee 100, 38116 Braunschweig, Germany}

\date{\today}
\begin{abstract}
	
	Extending quantum control to increasingly complex systems is crucial for both advancing quantum technologies and fundamental physics.
	In trapped ion systems, quantum logic techniques that combine a well-controlled logic species with a more complex spectroscopy species have proven to be a powerful tool for extending the range of accessible species.
	Here, we demonstrate that a quantum system as complex as \Ti with its many metastable states can be controlled employing a combination of intrinsic thermalization due to collisions with background gas and quantum-logic techniques using a far-detuned Raman laser. 
    The preparation of pure quantum states allows coherent manipulation and high resolution measurements of the Zeeman structure in \Ti. The presented techniques are applicable to a wide range of ionic species giving access to a larger variety of systems for fundamental physics and constitute the first step for quantum-controlled spectroscopy of transition metals, relevant, e.g., for the interpretation of astrophysical spectra.

\end{abstract}
\maketitle

Trapped ions are among the leading platforms for quantum technology~\cite{loschnauer_scalable_2024, acin_quantum_2018} and time and frequency metrology~\cite{ludlow_optical_2015}.  In group II ions and ions with similar structure, the exquisite control provided by direct laser cooling gives full control over the internal and external degrees of freedom~\cite{leibfried_quantum_2003}. Quantum logic techniques~\cite{schmidt_spectroscopy_2005} have extended this control to group III ions such as \Al ~\cite{schmidt_spectroscopy_2005, brewer_27+_2019}, highly-charged ions~\cite{micke_coherent_2020,king_optical_2022} and molecular ions~\cite{wolf_non-destructive_2016, chou_preparation_2017, sinhal_quantum-nondemolition_2020,holzapfel_quantum_2024}. However, applying quantum control techniques to transition metal ions with their large number of long-lived states has remained elusive. The abundance of transition metals in astrophysical spectra makes them excellent candidates for the investigation of variation of fundamental constants with quasar absorption spectra~\cite{murphy_laboratory_2013,webb_searching_2024} or studies of astrophysical objects~\cite{li_theoretical_2020,li_multiconfiguration_2020,pickering_recent_2019}.

Titanium is a transition metal with three valence electrons and a partially filled 3d shell, resulting in a complex electronic level structure with in total 17 metastable fine structure components in the  energetically low-lying even parity states with a lifetime longer than one minute~\cite{li_theoretical_2020}. Here, we demonstrate that despite this complex level structure, a combination of intrinsic thermalization due to background gas collisions in an ultra-high vacuum environment and quantum logic techniques using a far-detuned Raman laser, allow preparing pure quantum states. 
The quantum logic-based schemes presented here provide fast electronic state detection, which enables state tracking under collisions with the residual background gas. The preparation of a pure quantum state and state detection allows demonstrating coherent manipulation and high resolution spectroscopy of the Zeeman states by radio-frequency (rf) pulses. The presented techniques are applicable to a large variety of ionic species and have applications in radiative lifetime measurements~\cite{wood_improved_2013} as well as in the analysis of inelastic collisions~\cite{tayal_transition_2020}. Furthermore, it paves the way for the measurements of Land\'e $g$-factors with astrophysical relevance~\cite{li_multiconfiguration_2020} and precision spectroscopy in other transition metals with astrophysical relevance, exceeding the accuracy achieved in state-of-the-art collinear laser spectroscopy~\cite{ratajczyk_transition_2024}.%

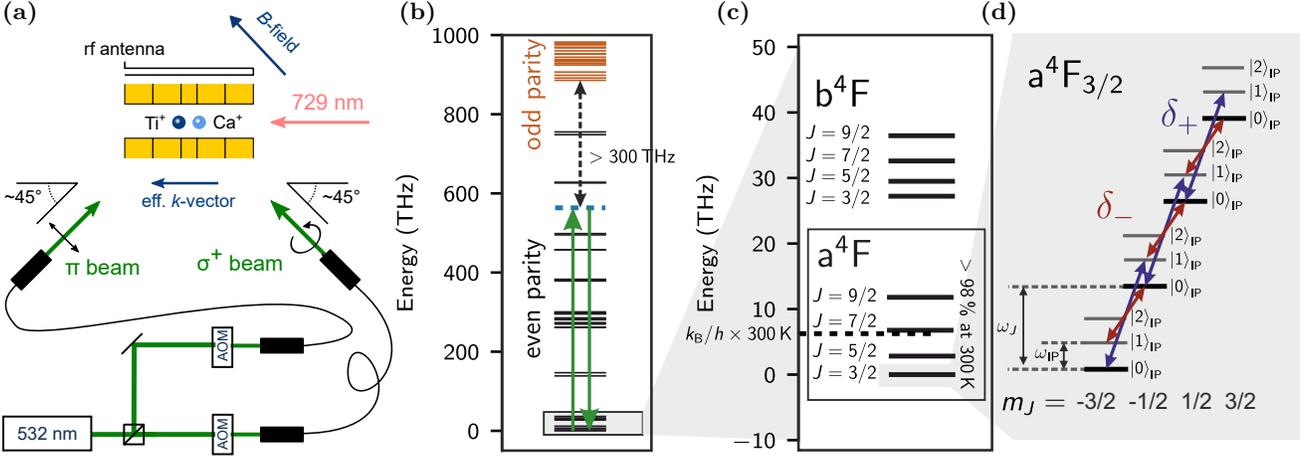
\begin{figure*}
    \begin{picture}(\textwidth,170)
        \put(0,165){\textbf{(a)}}
	    \put(0,0){\includegraphics[width=0.6\columnwidth]{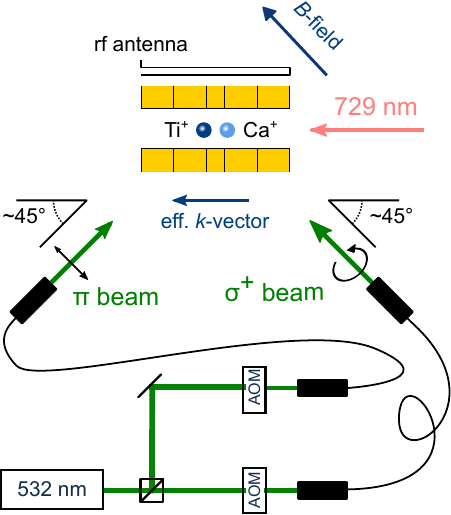}}
        \put(150,165){\textbf{(b)}}
        \put(270,165){\textbf{(c)}}
        \put(370,165){\textbf{(d)}}
        \put(0.6\columnwidth,0){\includegraphics[width=1.4\columnwidth]{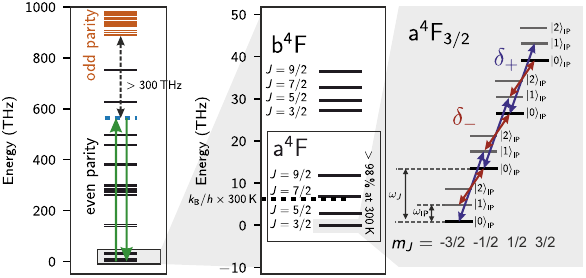}}
    \end{picture}
	\caption{ \textbf{(a)} Schematic of experimental setup with a segmented linear Paul trap holding a \Ca-\Ti ion crystal. Two Raman beams are formed from the same laser source, shifted in frequency by two acousto-optical modulators, and directed onto the two ions via optical fibers. They are $\pi$- and $\sigma^+$-polarized with respect to the quantization axis defined by the applied magnetic field. The alignment of the laser beams results in an effective wavevector $k$ along the trap axis. Carrier and sideband transitions between the $^2$S$_{1/2}$ and the $^2$D$_{5/2}$ levels of the logic ion are driven by a laser at \SI{729}{\nano\meter}. An in-vacuum rf antenna allows driving Zeeman transitions in \Ti. \textbf{(b)} Level structure of \Ti. Three valence electrons and a partially filled d-shell result in a complex level structure with many low-lying, long-lived even-parity states. The Raman laser (green arrows) couples to excited odd-parity states (orange) with a detuning larger than $\SI{300}{\tera\hertz}$. \textbf{(c)} In thermal equilibrium with the $\SI{300}{\kelvin}$ environment, more than $\SI{98}{\%}$ of the population is in the lowest-energy state \Fa, which has four fine structure components. \textbf{(d)} The magnetic field results in a fine-structure-dependent Zeeman splitting. Here, only the Zeeman splitting for the $J=3/2$ state is shown. The Raman laser is used with two different detunings $\delta_\pm$ to couple the Zeeman state $\ket{m_J}$ with the motional states $\ket{n}_\text{IP}$ of the axial in-phase~(IP) mode.}
	\label{fig:trap_setup}
\end{figure*}

  The experiments are performed in a segmented linear Paul trap \cite{leopold_cryogenic_2019} in a room-temperature ultra-high vacuum ($< \SI{1e-10}{mbar}$) system that is used to confine a two-ion ensemble consisting of a single \Ti and a single \Ca. 
  Following the approach of quantum logic spectroscopy, the co-trapped \Ca ion enables sympathetic cooling and state readout of the \Ti spectroscopy ion via shared motional modes due to the mutual Coulomb repulsion between the two ions. The axial motional mode frequencies of the coupled system are given by $\omega_{\text{IP}} = 2\pi \times \SI{660}{\kilo\hertz}$ and $\omega_{\text{OP}} = 2\pi \times \SI{1150}{\kilo\hertz}$ for the in-phase~(IP) and out-of-phase~(OP) mode, respectively. The two-ion crystal is prepared by loading multiple calcium and titanium ions from an ablation target~\cite{zimmermann_laser_2012} and subsequently reducing their number deterministically~\cite{zawierucha_deterministic_2024}.
  The isotope \Ti has been chosen due to its high natural abundance of $73.7\,\%$ \cite{kinsey_nudatpcnudat_2024} and the absence of hyperfine structure. Ablated calcium is isotope selectively photo-ionized~\cite{lucas_isotope-selective_2004}. Titanium is directly loaded from an ablation plasma without photoionization lasers. A magnetic field of $\SI{0.396}{\milli\tesla}$ along the \textit{z}-direction defines the quantization axis.
  For logic operations between the $\left|\downarrow \right> = \left| S_{1/2}, m_J=1/2 \right>$ and the $\left|\uparrow \right> = \left| D_{5/2}, m_J=-1/2 \right>$ qubit states in \Ca, we employ a narrow-linewidth laser at \SI{729}{\nano\meter} locked to a high finesse cavity for pre-stabilization. For long-term stability the laser is further stabilized to  an ultra-stable silicon cavity~\cite{matei_1.5_2017} using an optical frequency comb \cite{scharnhorst_high-bandwidth_2015}. The logic laser beam is oriented along the trap axis (see \cref{fig:trap_setup}~a) to address axial motional modes only.

Manipulation of the internal states in the \Ti spectroscopy ion is achieved using a far-detuned Raman laser setup  at \SI{532}{\nano\meter}. Its large detuning of $>\SI{300}{\tera\hertz}$ with respect to the closest dipole-allowed transition (see \cref{fig:trap_setup}~b) strongly suppresses off-resonant photon scattering~\cite{moore_photon_2023}, which would lead to state diffusion. This approached has already been demonstrated for molecular ions~\cite{chou_preparation_2017} and works as follows.

A single laser source provides two phase-coherent beams with \SI{\sim 100}{\milli\watt} of power.
These Raman beams are aligned as seen in \cref{fig:trap_setup}~a, which results in an effective wavevector $k$ along the axial direction, enabling sideband transitions and therefore coupling to the shared motional states. Choosing $\sigma^+$- and $\pi$-polarization for the two beams allows coupling of adjacent Zeeman states with $\Delta m_J=\pm1$. The relative angular frequency detuning $\delta = \omega_\sigma - \omega_\pi$ is controlled by acousto optical modulators (AOMs) in single-pass configuration. To achieve a high stability in the position of the beam foci, both beam paths are fiber coupled after passing the AOMs. Alignment of the two beams is achieved by maximizing the light shift on the logic transition of a single \Ca using piezo-actuated mirrors. The beam waist of around $\SI{30}{\micro\meter}$ ensures homogeneous illumination of a two-ion crystal. %
Tensor light shifts, which would result in a non-equidistant spacing between adjacent Zeeman states, are suppressed by tuning the intensity-ratio of the two Raman beams to  $I_{\sigma}/I_{\pi} = 2$ (see appendix \ref{app:tensor_shift_supp}).

Assuming a Boltzmann distribution, we expect that after thermalization with the room temperature environment by background gas collisions $>98\%$ of the state population of \Ti is in the a$^4$F ground state with electronic configuration 3d$^2$4s (see \cref{fig:trap_setup}~c). The four mainly populated fine structure states $\ket{\text{a}^4\text{F},J}$ can be distinguished by their Zeeman splitting $\omega_J$~\cite{loh_zeeman-splitting-assisted_2014,wolf_prospect_2024}. In the following, the individual sub-states $\ket{\text{a}^4\text{F},J, m_J}$ are denoted by $\ket{m_J}$ for brevity.

\begin{figure}
    \flushleft
    {\normalfont \normalsize \textbf{(a)} }
	\begin{scriptsize}
		\sffamily
		\def\svgwidth{\columnwidth}
		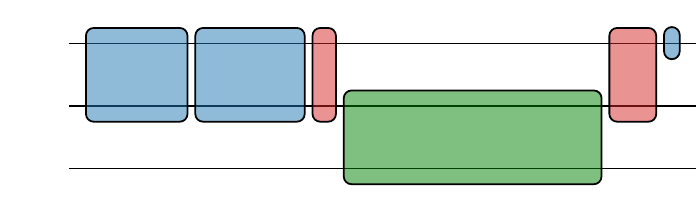
        {\normalfont \normalsize \textbf{(b)} }
        \includegraphics[width=\columnwidth]{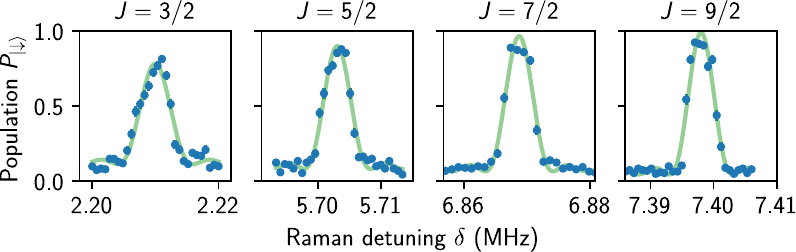}

		\end{scriptsize}

	\caption{Fine structure state detection in the \Ti a$^4$F ground state. \textbf{(a)} An experimental cycle starts with ground state cooling (GSC) composed of Doppler cooling (DC), electromagnetically-induced transparency cooling (EIT) and additional sideband cooling (SBC) cycles. A Raman pulse at \SI{532}{\nano\meter} and at alternating frequencies detunings $\delta_\pm$ transfers population between Zeeman states by simultaneously inducing motion to the ion crystal. State readout on the calcium ion in performed by a rapid-adiabatic-passage (RAP) RSB pulse on the in-phase mode followed by state detection (SD). \textbf{(b)} Measured Zeeman splitting for the $J \in \left\{3/2, 5/2, 7/2, 9/2\right\}$ states.}
	\label{fig:ti_transitions_lattice}
\end{figure}

The experimental sequence for fine structure state detection by measuring the Zeeman splitting is shown in \cref{fig:ti_transitions_lattice}~a. The state detection starts with ground-state cooling (GSC) of both axial modes by a sequence of Doppler and electromagnetically-induced transparency (EIT) cooling \cite{morigi_ground_2000} on the calcium ion. We perform additional resolved sideband cooling pulses alternating on the axial out-of-phase and in-phase mode of the coupled system. To detect the fine structure state $\ket{\text{a}^4\text{F}, J}$ in \Ti, GSC is followed by a sideband pulse on titanium that couples the states $\ket{m_J}\ket{n}_{\text{IP}}$ with $\ket{m_J+1}\ket{n\pm 1}_{\text{IP}}$ where $\ket{n}_\text{IP}$ denotes the axial in-phase motional state. As shown in \cref{fig:trap_setup}~d, this coupling is achieved by choosing a relative detuning $\delta$ between the $\pi$- and $\sigma^+$-polarized Raman beams of $\delta_\pm = \omega_J \pm \omega_\text{IP} + \omega_\text{VLS}$ with $\omega_J$ the Zeeman splitting for the state $\ket{\text{a}^4\text{F},J}$ and $\omega_\text{VLS}$ the vector light shift caused by the $\sigma^+$-polarized beam (see appendix \ref{app:tensor_shift_supp}). For the system initially being in the motional ground state, this interaction leads to motional excitation only if the \Ti is in the probed fine structure state. %
This motional excitation is detected on \Ca by a rapid-adiabatic passage pulse with the \SI{729}{\nano\meter} laser on the red sideband (RSB) transition between the two qubit states~\cite{gebert_detection_2016}. %
Initially, the titanium ion is in a mixture of $\ket{m_J}$ states. Repeated application of the state detection sequence, coupling adjacent Zeeman states by a relative Raman laser detuning of only $\delta_+$ or $\delta_-$ would eventually lead to accumulation of population in the edge state $\ket{m_J=+J}$ or $\ket{m_J=-J}$, respectively. To avoid pumping into this state and thereby producing a dark state, detuning is switched between $\delta_+$ and $\delta_-$ for each experimental cycle.
We also use RAP for the Raman interaction by sweeping $\delta$ over \SI{15}{\kilo\hertz} during each Raman pulse. This compensates residual tensor shifts and accounts for different Rabi frequencies for the involved $\ket{m_J}\ket{n}_\text{IP}$ states. 

\begin{figure}
        \flushleft
        \textbf{(a)}
        \includegraphics[width=\columnwidth]{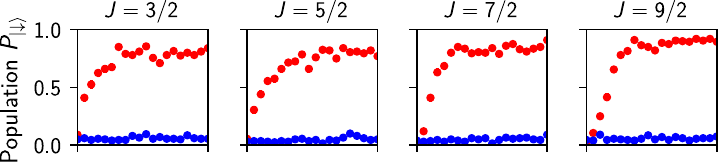}
        \\\vspace*{10pt}
        \textbf{(b)}
        \includegraphics[width=\columnwidth]{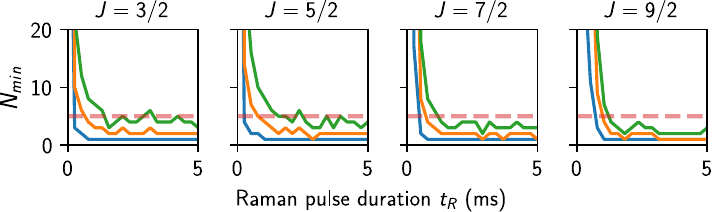}
	\caption{Evaluation of the fine structure state detection efficiency. \textbf{(a)} Excitation probability on the logic ion in dependence of the Raman pulse length $t_\text{R}$ when \Ti is in the given state $\ket{J}$ (red) or when it is in one of the other states (blue). Each point is an average over $N=200$ measurements. \textbf{(b)} Minimum number of experimental cycles ($N_\text{min}$) required to distinguish whether the titanium ion is in a probed state or not. $N_\text{min}$ is calculated for detection confidence (see \cref{eq:min_exp_cycles}) for $1\sigma_c$  (blue), $2\sigma_c$ (orange), and $3\sigma_c$ (green). See text for details. The red dashed line is at $N_\text{min}=5$.}
	\label{fig:ti_state_detec_bench}
\end{figure}

\begin{figure}
\begin{picture}(\columnwidth,185)
	  \put(0,95){\includegraphics[width=\columnwidth]{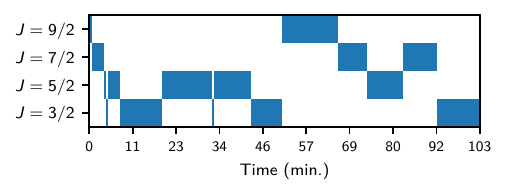}}
   \put(0,0){\includegraphics[width=\columnwidth]{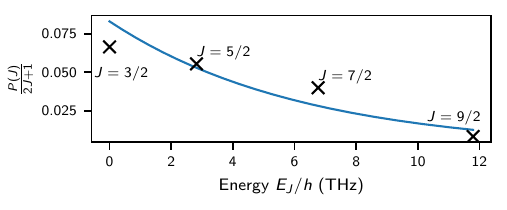}}
   \put(0,185){\textbf{(a)}}
   \put(0,100){\textbf{(b)}}
      
\end{picture}	
 \caption{\textbf{(a)} Real-time state tracking of \Ti by continuous application of the state detection sequence described in the main text. Blue bars indicates occupation of the respective fine structure state for the given time period. \textbf{(b)} Black crosses show the measured proportion of the population $P(J)$ in state $J$, inversely weighted by the degeneracy factor $2J+1$ for a measurement time of around $\SI{16}{\hour}$. The blue line shows a theoretical Boltzmann distributions, which is expected for thermal equilibrium in a $\SI{300}{\kelvin}$ environment (blue bars/solid line). }
	\label{fig:ti_state_tracking}
\end{figure}

Guided by theoretical predictions for the \textit{g}-factor of the respective transitions~\cite{li_multiconfiguration_2020}, four different states with total angular momentum $J\in\left\{3/2, 5/2, 7/2, 9/2\right\}$  have been detected. Scans over the four Zeeman frequencies $\omega_J$, using the described detection method, are shown in \cref{fig:ti_transitions_lattice}~b.
For state detection, these four peaks are probed on resonance and the logic ion excitation $P_s$ is compared to the background excitation $P_b$, which is the excitation measured when the \Ti is not in the probed state. 
The signal-to-noise ratio for state detection is therefore $ n_\text{SNR} = |P_s-P_b|/\sigma_c$, with the combined quantum projection noise~\cite{itano_quantum_1993} $\sigma_c = \sqrt{\sigma_s^2+ \sigma_b^2}$, where $\sigma_{s/b}^2 = P_{s/b}(1-P_{s/b})/N$ and the number of repetitions of the detection sequence, $N$. Thus, we can define the minimum number of experimental repetitions $N_\text{min}$ required to achieve a signal-to-noise ratio~(SNR) of $R$ by
\begin{equation}
	\label{eq:min_exp_cycles}
	N_\text{min} = \left\lceil\frac{P_s(1-P_s) + P_b(1-P_b)}{\left|P_s - P_b\right|}R^2\right\rceil.
\end{equation}
\cref{fig:ti_state_detec_bench}~a shows the excitation and background signal for \Ti state detection for different Raman laser detection pulse durations $t_\text{R}$. In \cref{fig:ti_state_detec_bench}~b, $N_\text{min}$ is inferred from this data. It can be seen that 5 repetitions of the detection sequence with a pulse duration of $t_\text{R} \geq \SI{2.5}{\milli\second}$ are sufficient to detect the \Ti fine structure state with a confidence larger than $3\sigma_c$.
In consequence, state detection with a temporal resolution of approximately \SI{80}{\milli\second} can be achieved, including also the time required for operations on the logic ion.
This feature allows tracking the quantum state of titanium in real time and observing transitions between the fine structure states. The implemented sequence probes all four states sequentially. The state with the highest excitation probability is identified as the actual fine structure state of the titanium ion. A typical time trace for the state evolution of a single \Ti is shown in \cref{fig:ti_state_tracking}~a. Collected data over $\SI{16}{\hour}$ in total show an average transition rate of \SI{7.4}{\per\hour}. Since radiative life times for the investigated states are larger than \SI{150}{\min} the majority of the state transitions can be attributed to collisions with residual background gas. Comparing the state distribution of the recorded data with a Boltzmann distribution at $\SI{300}{\K}$ confirms the assumption of thermalization with the room temperature environment (see \cref{fig:ti_state_tracking}~b). We attribute the residual discrepancy to statistical fluctuations due to the finite sampling time. %


\begin{figure}
        \flushleft
        {\normalfont \normalsize \textbf{(a)} }\\\vspace*{-0.1cm}
        \begin{scriptsize}
		\sffamily
		\def\svgwidth{\columnwidth}
		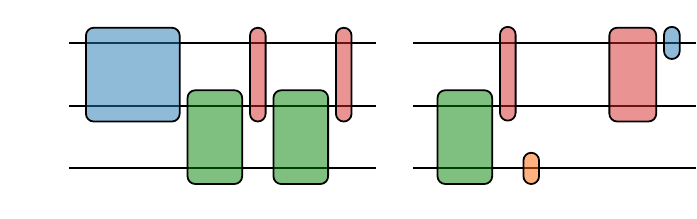
  		\vspace*{10pt}
        {\normalfont \normalsize \textbf{(b)} }\\\vspace*{-0.7cm}
		\includegraphics[width=\columnwidth]{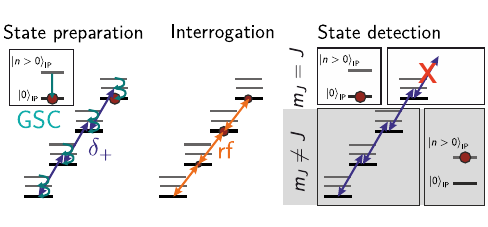}
        {\normalfont \normalsize \textbf{(c)} }\\
        \includegraphics[width=\columnwidth]{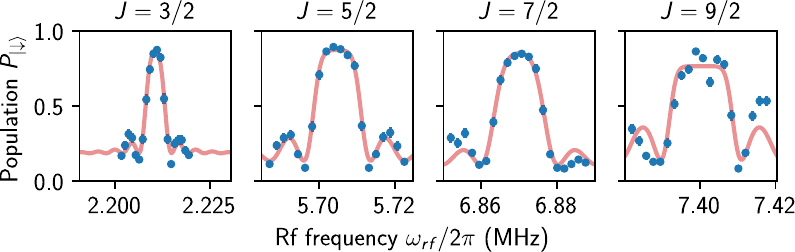}

	\end{scriptsize}
	\caption{Measurement of the Zeeman splitting in the \Ti ground state using radio frequency (rf). \textbf{(a)} Experimental cycle starting with ground state cooling (GSC) (see \cref{fig:ti_transitions_lattice}a). Dissipative pumping to edge state of Zeeman manifold by twenty repetitions of a Raman pulse at frequency detuning $\delta_+$ followed by twenty sideband cooling pulses to dissipate motion. An rf pulse is applied to depopulate the edge state again, when $\omega_\text{rf}=\omega_{J}$. An additional Raman pulse at $\delta_{+}$ induces motion only when depopulation of the edge state was successful. The sequence ends by state readout on the logic ion via a RAP pulse and fluorescence detection. \textbf{(b)} Spectroscopy sequence illustrated on the $J=3/2$ Zeeman states. See text for details. \textbf{(c)} Fourier-limited rf spectra of Zeeman transitions for $J \in \left\{3/2, 5/2, 7/2, 9/2\right\}$.}
	\label{fig:ti_transitions_rf}
\end{figure}

\begin{figure}
	\includegraphics[width=\columnwidth]{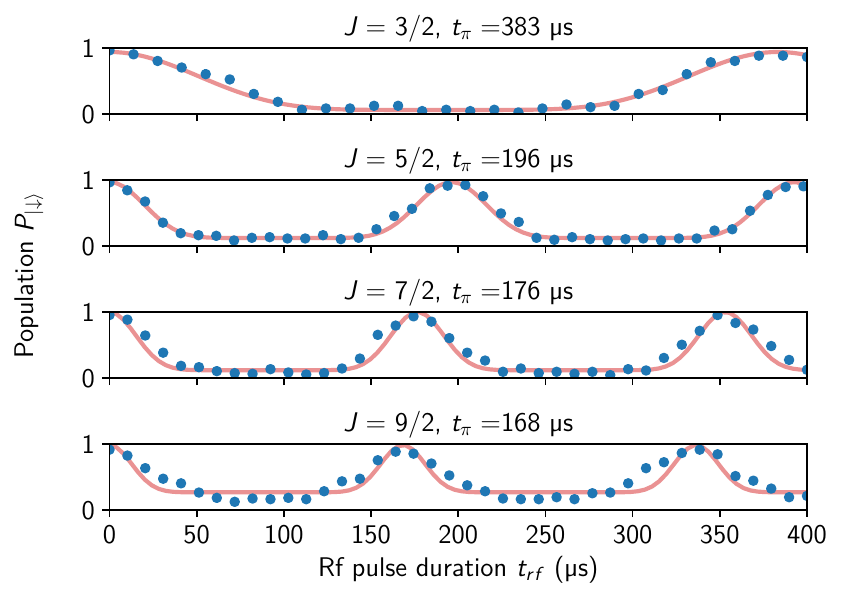}
	\caption{Radio frequency Rabi oscillation for different fine structure states $J \in \left\{3/2, 5/2, 7/2, 9/2\right\}$. The excitation of the logic ion qubit indicates the probability to find \Ti in the edge state $\left|J, m_J=J\right>$ after initial preparation and rf interaction time $t_\text{rf}$. Equation (\ref{eq:rabi_rf}) was used to fit the data and extract individual times $t_\pi$.}
	\label{fig:ti_rf_rabi_flopping}
\end{figure}

Successful state detection is a prerequisite for an extended protocol that has been implemented to prepare a pure quantum state of the titanium ion.
For this, ground state cooling of the two-ion crystal is followed by a sequence of sideband pulses with detuning $\delta_+$ with the Raman laser adding a quantum of motion for each increase in $m_J$. These pulses are interleaved with ground state cooling on the logic ion providing a dissipation channel. Eventually, this pulse sequence leads to accumulation of population in the edge state $\ket{m_J=+J}$~\cite{schmidt_spectroscopy_2005, chou_preparation_2017}.
Lack of excitation of the motional mode indicates successful preparation of the edge state (see \cref{fig:ti_transitions_rf}~b right panel)
Preparing pure quantum states with high fidelity enables coherent control and high-resolution spectroscopy. This is showcased by rf-induced Zeeman transitions using an in-vacuum antenna.
Once the edge state is prepared, an rf pulse at frequency $\omega_{rf} = \omega_{J}$ is applied, which implements a global spin rotation of the $\ket{m_J}$ by an angle $\theta$ resulting in the state $\ket{\psi}=\exp(i/\hbar \hat J_x \theta)\ket{m_J=J}$, where $\hat J_x$ is the projection of the angular momentum operator in the $x$-direction.
A subsequent sideband transition using the Raman laser and detection of motional excitation provides a measurement of $|\braket{m_J=J|\psi}|^2$ (see figure \cref{fig:ti_transitions_rf}~b). The result of such an experiment for different rf pulse durations $t_\text{rf}$ is shown in \cref{fig:ti_rf_rabi_flopping} for all four fine structure states in a$^4$F. 
The data allows inferring the time $t_\pi$ required for rotating the spin by an angle $\theta=\pi$~\cite{radcliffe_properties_1971} using the fit function 
 \begin{equation}
 	\label{eq:rabi_rf}
 	P_{\ket{m_J=J}}(t_\text{rf}) = a \left(1-\cos\left(\pi\cdot \frac{t_\text{rf}}{t_{\pi}}\right)^{4J}\right) + c,
 \end{equation} with free parameters a, c and $t_\pi$.
  The sequence also allows high resolution Rabi spectroscopy with an rf pulse duration of $t_\text{rf} = t_{\pi}/2$ on the Zeeman transitions, as illustrated by the Fourier-limited spectra shown in \cref{fig:ti_transitions_rf}~c.

We have demonstrated that collision-induced thermalization in combination with quantum-logic techniques provides extensive quantum control capabilities for a system as complex as \Ti. The technique can be applied to other transition metals and beyond, making an entire new class of ions accessible to quantum logic spectroscopy, which was previously considered infeasible due to the high number of metastable states. For instance, $^{56}$Fe$^+$ is of exceptional importance in astrophysics. Due to the large fine structure splitting in the a$^6$D ground state, approximately $80\,\%$ of the population should occupy the $J=9/2$ state after thermalization at \SI{300}{\kelvin}. This allows for quantum state preparation with a high duty cycle using the methods presented here. The high resolution achieved for rf spectroscopy allows measuring Land\'e $g$-factors with high precision. The co-trapped logic ion can serve as a precise sensor for calibrating magnetic fields~\cite{sailer_measurement_2022}.
Precision measurements of the \Ti Land\'e $g$-factor using Ramsey spectroscopy are currently ongoing. 
The real-time tracking and detection of fine structure state changing collisions provides a new tool to investigate inelastic collisions between single ions and neutral background gas, which is an important cooling mechanism in interstellar media~\cite{tayal_transition_2020,goldsmith_collisional_2012,wiesenfeld_c_2013}.

\section*{Acknowledgments}
We thank Alexander Wilzewski for helpful comments on the manuscript. Furthermore, we thank the iqloc team for providing us with pre-stabilized light for $\SI{729}{\nano\meter}$.
This research was funded by the Deutsche Forschungsgemeinschaft (DFG, German Research Foundation) – Project-ID 274200144 – SFB 1227 (DQ-mat) B05 with partial support from Germanys	Excellence Strategy EXC-2123 QuantumFrontiers 390837967. The project has received funding
from the European Research Council (ERC) under the European Union’s Horizon 2020 research and innovation programme (grant agreement No 101019987)

\appendix

\section{Tensor light shift compensation}
\label{app:tensor_shift_supp}
The ac-Stark shift $\Delta E_\text{LS}$ of a state $\ket{J,m_J}$ from a light field with electric field $\vec \varepsilon=(\varepsilon_x,\varepsilon_y,\varepsilon,_z)$ can be expressed as~\cite{stalnaker_dynamic_2006}
\begin{align}
    \begin{split}
        \Delta E_\text{LS} = &-\frac{\alpha_0}{2}|\vec\varepsilon|^2\\&- \underbrace{\frac{1}{2}\frac{m_J}{J} V_z}_{\Delta E_{VLS}}\\& -\underbrace{\frac{\alpha_2}{2}\left( \frac{3m_J^2-J(J+1)}{J(2J-1)}\right)\frac{3|\varepsilon_z|^2-|\vec\varepsilon|^2}{2}}_{\Delta E_\text{TLS}}
    \end{split}
\end{align}

with $V_z=\frac{1}{2}\left(\varepsilon_x\varepsilon_y^*-\varepsilon_x^*\varepsilon_y\right)$, the $z$-component of the circular intensity vector and the scalar, vector and tensor polarizabilities $\alpha_0$, $\alpha_1$ and $\alpha_2$, respectively.
The dependence of the individual terms on the angular momentum projection quantum number $m_J$ indicates, that the scalar part results in a constant offset of all Zeeman components, the vector part ($\Delta E_\text{VLS} = \hbar\omega_\text{VLS}$) resembles an additional magnetic field and only the tensor part results in a non-equidistant spacing between adjacent Zeeman states. This non-equidistant spacing presents an obstacle for the demonstrated state detection and preparation protocols.  
Appropriate choices of polarization suppress the tensor light shift~\cite{wineland_quantum_2003,chou_preparation_2017}. The tensor light shift $\Delta E_\text{TLS}$ for the $\pi$-polarized Raman beam (i.e. $|\vec\varepsilon^\pi|=|\varepsilon^\pi_z|$) is given by 
\begin{align}
    \Delta E^\pi_\text{TLS} =  \frac{\alpha_2}{2}\left( \frac{3m_J^2-J(J+1)}{J(2J-1)}\right)\frac{2|\vec\varepsilon^\pi|^2}{2},
\end{align}
and for $\sigma^+$ polarized light (i.e. $\varepsilon^\sigma_z=0$) by 
\begin{equation}
    \Delta E^\pi_\text{TLS} =  -\frac{\alpha_2}{2}\left( \frac{3m_J^2-J(J+1)}{J(2J-1)}\right)\frac{|\vec\varepsilon^\sigma|^2}{2}.
\end{equation}
Therefore, the total tensor shift for both light fields interacting with the ion is 
\begin{equation}
 \Delta E^\pi_\text{TLS} +\Delta E^\pi_\text{TLS} \propto 2|\vec\varepsilon^\pi|^2-|\vec\varepsilon^\sigma|^2   
\end{equation}
and can be suppressed by  operating with an intensity ratio of $I_{\sigma}/I_{\pi} = 2$. 

\bibliographystyle{apsrev4-2}

\bibliography{ti-qls_bib}

\end{document}